\documentclass[sigconf]{acmart}

\AtBeginDocument{%
  \providecommand\BibTeX{{%
    \normalfont B\kern-0.5em{\scshape i\kern-0.25em b}\kern-0.8em\TeX}}}

\setcopyright{acmcopyright}
\copyrightyear{2018}
\acmYear{2018}
\acmDOI{XXXXXXX.XXXXXXX}

\acmConference[Conference acronym 'XX]{Make sure to enter the correct
  conference title from your rights confirmation emai}{June 03--05,
  2018}{Woodstock, NY}
\acmPrice{15.00}
\acmISBN{978-1-4503-XXXX-X/18/06}

\copyrightyear{2023} 
\acmYear{2023} 
\setcopyright{rightsretained} 
\acmConference[ICONS '23]{International Conference on Neuromorphic Systems}{August 1--3, 2023}{Santa Fe, NM, USA}
\acmBooktitle{International Conference on Neuromorphic Systems (ICONS '23), August 1--3, 2023, Santa Fe, NM, USA}
\acmDOI{10.1145/3589737.3605977}
\acmISBN{979-8-4007-0175-7/23/08}

\begin{document}
\title{Dendritic Computation through Exploiting Resistive Memory as both Delays and Weights}


\author{Melika Payvand}
\affiliation{%
  \institution{Institute of Neuroinformatics, University of Zurich and ETH Zurich}
  \streetaddress{Wintherthurerstrasse 190}
  \city{Zurich}
  \country{Switzerland}}
\email{melika@ini.uzh.ch}

\author{Simone D'Agostino}
\affiliation{%
  \institution{Institute of Neuroinformatics, University of Zurich and ETH Zurich}
  \city{Zurich}
  \country{Switzerland}
}

\author{Filippo Moro}
\affiliation{%
 \institution{CEA-LETI}
 \city{Grenoble}
 \country{France}}

\author{Yigit Demirag}
\affiliation{%
 \institution{Institute of Neuroinformatics, University of Zurich and ETH Zurich}
 \city{Zurich}
 \country{Switzerland}}

\author{Giacomo Indiveri}
\affiliation{%
  \institution{Institute of Neuroinformatics, University of Zurich and ETH Zurich}
  \city{Zurich}
  \country{Switzerland}}

\author{Elisa Vianello}
\affiliation{%
  \institution{CEA-LETI}
  \city{Grenoble}
  \country{France}}

\renewcommand{\shortauthors}{Payvand, et al.}

\begin{abstract}
  Biological neurons can detect complex spatio-temporal features in spiking patterns via their synapses spread across their dendritic branches. This is achieved by modulating the efficacy of the individual synapses, and by exploiting the temporal delays of their response to input spikes, depending on their position on the dendrite. Inspired by this mechanism, we propose a neuromorphic hardware architecture equipped with multiscale dendrites, each of which has synapses with tunable weight and delay elements. Weights and delays are both implemented using Resistive Random Access Memory (RRAM). We exploit the variability in the high resistance state of RRAM to implement a distribution of delays in the millisecond range for enabling spatio-temporal detection of sensory signals. We demonstrate the validity of the approach followed with a RRAM-aware simulation of a heartbeat anomaly detection task. In particular we show that, by incorporating delays directly into the network, the network's power and memory footprint can be reduced by up to 100x compared to equivalent state-of-the-art spiking recurrent networks with no delays. 
\end{abstract}




\newcommand{\MP}[1]{\textcolor{red}{{\bf MP:}~#1}} 
\newcommand{\SD}[1]{\textcolor{cyan}{{\bf SD:}~#1}} 


\maketitle

\section{Introduction}

\begin{figure}[ht]
  \centering
  \includegraphics[width=0.7\linewidth]{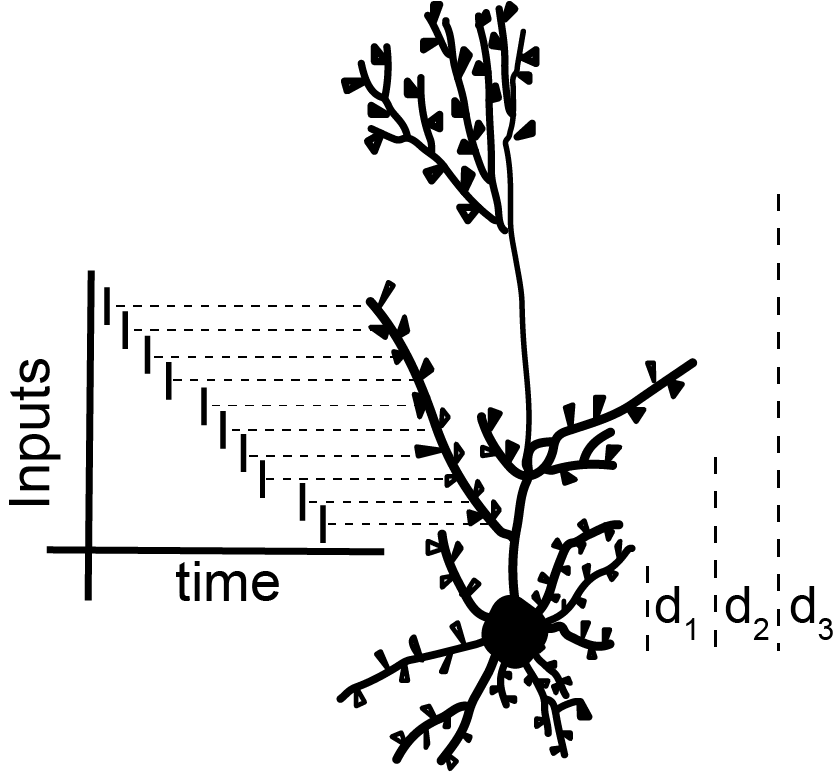}
  \caption{Biological neurons include synapses distributed spatially in their dendritic arbor, which gives rise to delayed inputs. The coincidence of the delayed spikes are detected as the input features in each dendrites. $d_1$, $d_2$ and $d_3$ show the average delay of each dendritic compartment depending on their spatial arrangement with respect to the neuron's cell body (soma). 
}
\label{fig:bio_den}
\end{figure}

In the typical artificial neural network models, the neuron's output is a nonlinear transformation of the weighted sum of its inputs, using a point-neuron model. In point neuron models all synapses are connected to the same node and their spatial position carries no extra information. Although for static rate-based information encoding, point-neuron models have enough complexity to perform computation, they are not ideal for detecting the temporal aspects of dynamic input patterns. Neuroscience findings show that the dendritic arbor of a neuron implements non-linear integration in multiple time-scales, and decodes spatio-temporal locality of arriving events, a mechanism known as coincidence detection (CD)~\cite{paugam_etal_2012_SNN,pagkalos_etal2023_dendrify}.
CD is highly dependent on the spatial arrangement of the synapses on the dendrites, which affects the timing of the arrival of the input spike to the neuron's soma~\cite{pagkalos_etal2023_dendrify} (Fig.~\ref{fig:bio_den}). This spatial arrangement can be modeled as synaptic delays, serving as an additional parameter for synapses alongside their weight.. In this sense, each synapse can be modeled as a combination of a temporal variable (delay) and a spatial variable (weight). It has already been shown that training temporal variables such as adaptation time constant of the neuron can improve the accuracy of Spiking Neural Networks (SNNs) in classifying spatio-temporal patterns~\cite{yin_etal2021_SRNNAdapt}. Similarly, endowing silicon neurons with dendritic circuits enables them to detect spatio-temporal patterns~\cite{Wang_Liu10}.
However, so far, no hardware implementation of spiking neural networks where dendritic temporal delays are learnt has been proposed.
Here, we propose an event-based architecture based on Resistive Random Access Memory (RRAM) that implements both delays and weights.
We exploit the strong programming variability of the HfO$_2$-based RRAM in its High Resistive State (HRS) to sample synaptic delays from the range of milliseconds, and program RRAM devices to tune synaptic weights. We show that our approach enables more efficient processing of spatio-temporal sensory signals in real-time, using only feed-forward networks, without resorting to recurrency, and demonstrate how it reduces the memory and power footprint by two orders of magnitude, compared to equivalent Recurrent Neural Networks (RNNs).

\section{RRAM-based dendritic computation}

\begin{figure*}[t]
  \centering
  \includegraphics[width=0.9\linewidth]{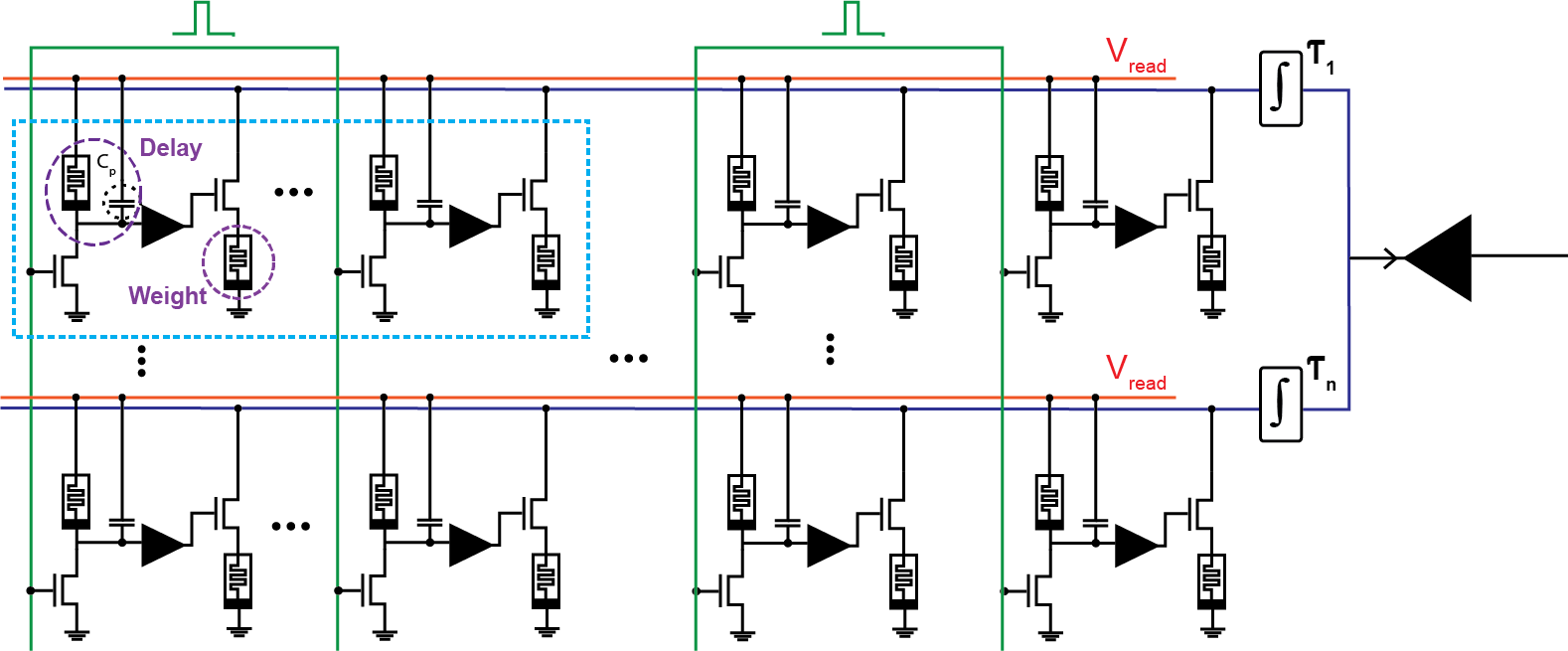}
  \caption{ Dendritic architecture using complex synapses containing RRAM delays and RRAM weights. Each channel (shown in green) is applied to a parallel set of synapses (in dashed blue box) in each row, which constitutes a distribution of delays, of which a sample is taken through learning the weight values. Each branch/row can integrate the delayed and weighted input channels with a different time constant $\tau_i$. 
}
  \label{fig:rram_den}
\end{figure*}

Inspired by the dendritic structure of the biological neurons of Fig. ~\ref{fig:bio_den}, we propose a hardware architecture equipped with multiscale dendrites, each of which has synapses with tunable weight and delay elements, implemented using RRAM (Fig. ~\ref{fig:rram_den}). The delay is implemented using an RRAM coupled with a capacitor (the RRAM-C element), while the weight is represented by one RRAM device.  
A dendritic circuit is then constituted by a RRAM-C element, activated by input spikes applied to an access transistor, and by an output section featuring the weight RRAM, outputting a weighted current pulse.

Dendritic circuits can be arranged into arrays, as shown in Fig.~\ref{fig:rram_den}. Each row constitutes a dendritic branch, with synapses that have both delay and weight elements. The synaptic delays of each dendritic branch have a certain distribution with a mean that is different from other branches. The green columns receive the spatio-temporal inputs, and each column receives the input from a different channel. The input spikes from these channels go through delays and get weighted and are then filtered by a different time constant ($\tau_i$). The delayed, weighted and integrated current contributions  are then summed to the neuron's soma on the right end, which is modeled as Leaky integrate and fire (LIF).  
To learn to classify spatio-temporal signals in this architecture, each dendritic branch needs to detect signal features at its integration time scale, through coincidence detection. In other words, the delay and weight parameters should be configured to perform CD in the presence of an input feature.  This makes relevant spikes available to the output neuron with temporal coincidence and leads the output neuron to produce spikes in turn. 

\begin{figure}[ht]
  \centering
  \includegraphics[width=0.7\linewidth]{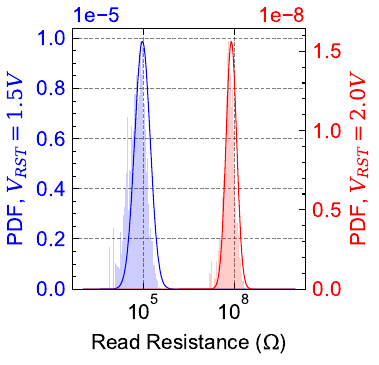}
  \caption{Variability of RRAMs in their HRS follows a wide log-normal distribution. The shift in the distribution is as a result of different reset voltages.
}
\label{fig:HRS_RRAM}
\end{figure}

To enable real-time processing, the delayed elements should be in the range of the time constant of the sensed real-world signals, e.g., in the order of 10s-100s of milliseconds. Thus, to implement such delays on-chip, while reducing the capacitor size, we exploit the HRS of RRAMs. Since the conductive filament resulting in resistive switching is very weak in the HRS, controlling the precise value of the resistance of RRAMs in the HRS is difficult. This can be seen in the HRS measurements preformed on HfO$_2$-based RRAM~\cite{esmanhotto_etal_2022} shown in Fig.~\ref{fig:HRS_RRAM}, with large variability in the HRS following a log-normal distribution. The mean of this distribution is a function of the reset voltage with which the device is switched to the HRS~\cite{Dalgaty_etal19}. 
Due to this variability, resetting the delay devices using the same voltage results in samples from the corresponding log-normal distribution. Each dendritic branch then features a variable delay with a certain mean, proportional to the mean HRS of the RRAM multiplied by the capacitance C. 
The network objective is then to learn the correct weights corresponding to the delay samples from this log-normal distribution, such that the neuron performs coincidence detection, reacting to the temporal features of the signal.

\section{RRAM-aware training}


The HRS of delay RRAMs cannot be precisely controlled. Therefore, prior to the training, we initialized the resistance values of delay RRAMs by sampling from HRS and kept them fixed. This substantial variability enables dendritic architecture to take advantage of a range of delay values.

The dendritic architecture poses some constraints in the offline training procedure, which have to be accounted for in order to extract its full potential. 
In the current configuration of the architecture, weight-RRAMs only express positive weights with limited precision (approximately 3 bits~\cite{esmanhotto_etal_2022}), contrary to the 32 bit floating point precision available on CPU/GPU. Also, the resistance is limited to a certain interval that delimits the Low-Resistive-State (LRS), which in our case spans from 7k$\Omega$ to 50k$\Omega$. The HRS can also be utilized in the weight-RRAMs when the algorithm selects low weights, although the LRS is preferable for weight-RRAMs as it is more controllable.
Moreover, the weight value in such devices is not deterministic \cite{esmanhotto_etal_2020}, i.e. the resistance value in LRS after the programming operation can be modeled as sampling from a Gaussian distribution whose mean is determined by the programming operation, and its standard deviation is due to the device non-idealities and cannot be controlled.

Due to the variability of RRAMs, offline training of the dendritic architecture has to be tailored to the RRAM characteristics.
In this work, a simple weight-clipping is used after the weight-update to ensure all weights remain positive and within the permitted range or resistance.

The limited precision is accounted for using a mixed-precision approach \cite{Dedemirag_etal_2021_mixedprecision,legalo_etal2018_mixprec}.
Gradients calculated with the backprogagation algorithm are accumulated on high-precision variables on an external computer.
At the end of each epoch, this variable is checked and - if the change passes the quantization step - the related RRAM device is reprogrammed. In such cases, the weight is updated by sampling its value from the new corresponding Gaussian distribution.

More precisely, the set of resistive levels assumed by the RRAM is defined by $\mu_n$, where $n$ goes from 1 to 8 (3 bits), each representing a resistance in the LRS. The high-precision variable (32-bit), also said hidden-weight $W_{ij}$, triggers a reprogramming operation when it approaches a new resistive level $\mu_j$, starting from a different value. 
\begin{equation}
    \mu:|W_{ij}-\mu|=\min_{k=0}^n\left\{|\mu_k-W{ij}|\right\}
    \label{eqn:1}
\end{equation}
where $n$ is the number of available resistive levels on the RRAM devices.
The RRAM-aware training procedure is summarized below: 
\begin{itemize}
    \item $n_{pre}$ epochs of pre-training on the 32-bit weights only, obtaining the pre-trained parameters $W_{pre}$;
    \item converting the hidden weights $W_{pre}$ to RRAM values after updating the scaling factor $s_w$ relating the resistance of the RRAM to the hidden weight; 
    \item $n_{training}$ epochs on the 3-bit precision RRAM weights, i.e. the values sampled from the LRS levels, scaled by $s_w$. 
\end{itemize}
with $s_w$ obtained as $\max{LRS}/\max{W_{trained}}$.

Importantly, the resistive levels $\mu_i$ and the standard deviation values related to the RRAM resistance are obtained from a 4kbit RRAM array operated with the smart programming procedure, as in \cite{esmanhotto_etal_2020}.

\section{Results}

\begin{figure*}[t]
  \centering
  \includegraphics[width=0.8\linewidth]{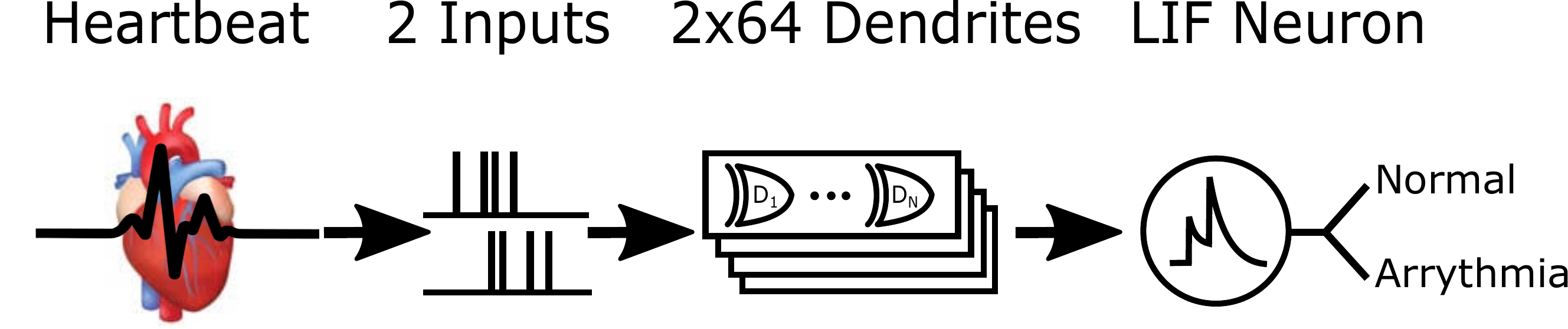}
  \caption{Arrhythmia detection with the dendritic architecture. The voltage recording of the heartbeat is converted to spike-trains and then fed to the Dendritic Architecture. An output neuron fires to signal the anomalies in the heartbeats.
}
\label{fig:ECG_task}
\end{figure*}

\begin{figure}[h]
    \centering
    \includegraphics[width=0.9\linewidth]{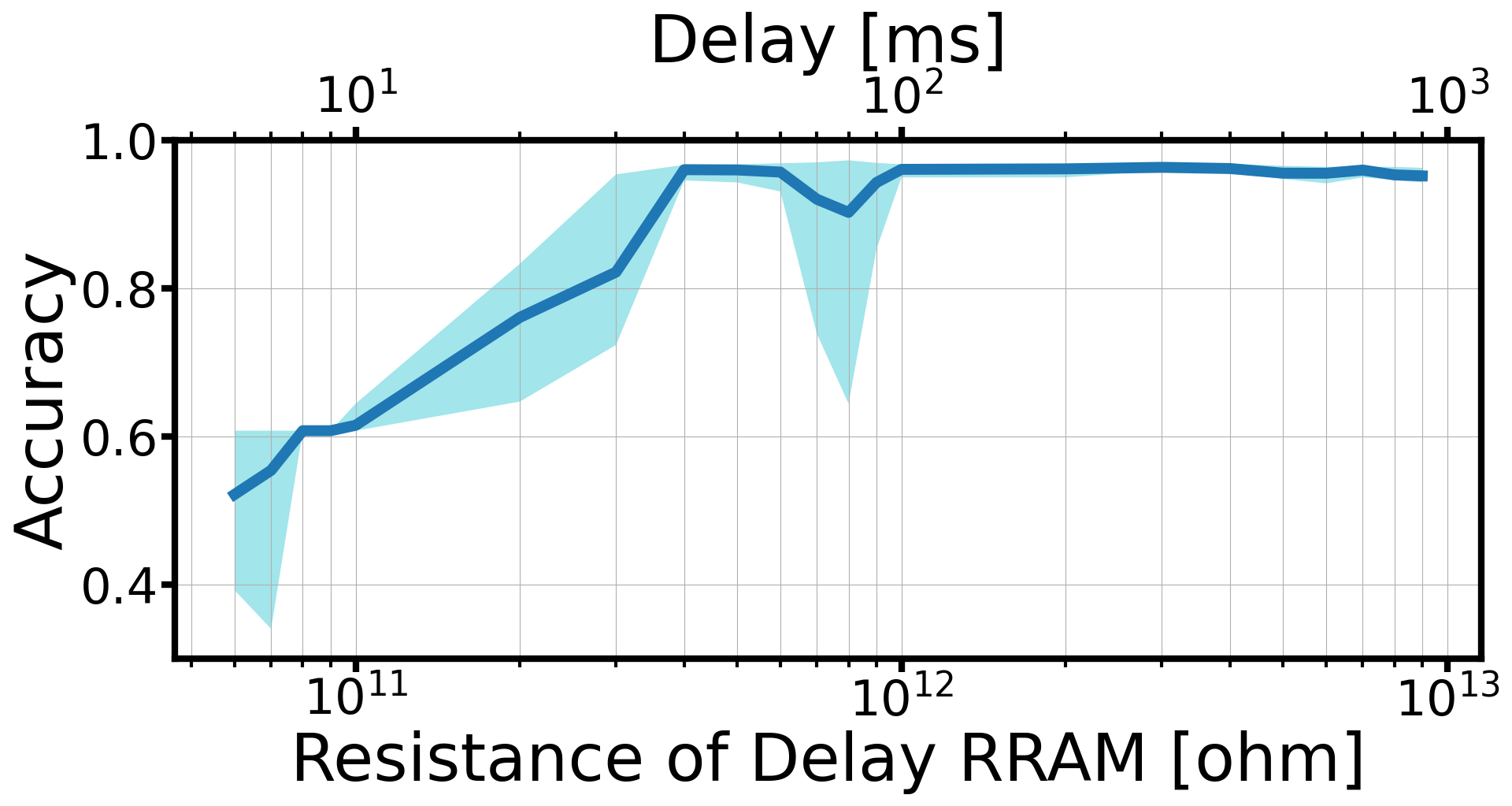}
    \caption{Accuracy of the Dendritic Architecture on the ECG arrhythmia detection, as a function of the delay-RRAM mean resistance.}
    \label{fig:acc_v_delay}
\end{figure}

To show-case the computational power of dendrites, we benchmark the architecture of Fig.~\ref{fig:rram_den} on a real-time sensory processing task, namely heartbeat anomaly detection, using Electrocardiogram (ECG) data. We choose the MIT-BIH dataset~\cite{moody_etal_2001_mitbih} and focus on the data of patient 208, presenting a balanced amount of normal and abnormal heartbeats. The raw data, consisting of the voltage traces recorded from different electrodes, is delta-modulated to obtain spike trains that are fed to the dendritic architecture~\cite{sharifshazileh_etal2021_eeg} (Fig~\ref{fig:ECG_task}).
The spiking activity of the output neurons signals the presence of arrhythmia in the heartbeat, performing binary classification. 
Importantly, the accuracy in solving this task depends on how well the temporal features in a heartbeat signal are interpreted to identify anomalies. In our particular model, the dendritic architecture, this means that the delay values have to match the temporal features of the input signal. The average heartbeat duration is on the order of $700$~ms, so the relevant temporal features should be a fraction of that period. These temporal features are detected through the delays. To find the average value of the delays required to detect the ECG features, we sweep the mean of the delay RRAM distribution, while fixing the capacitance size to $100~fF$. Figure~\ref{fig:acc_v_delay} shows the accuracy as a function of the mean value of the log-normal distribution related to the delay RRAM, with the equivalent delay shown on top of the figure. As can be seen, the task is solved (i.e. accuracy > 95\%) with a mean delay of $40~ms$. This delay corresponds to a HRS of 500 $GOhms$ which is difficult to achieve with HfO$_2$-based devices. However, their pristine state can be used to achieve this resistance. Alternatively, Ferroelectric Tunnel Junction devices are promising candidates for such large resistance levels~\cite{barbot2022interplay}.

Using the mean delay of 40\,ms, a single output neuron, with two dendritic branches of 64 synapses each, can achieve up to 95\% accuracy on the real-time ECG anomaly detection task. 
This is compared to more than 100 units required in Spiking Recurrent Neural Networks (SRNNs) from previous works, giving rise to 100~$\times$ reduction in power consumption for the aforementioned task~\cite{gutig_etal_2006_tempotron,Bauer_etal_2019_ECGDynap}. 
Table~\ref{tab:power} shows the comparison of the estimated power consumption and memory footprint of the dendritic architecture against other state of the art methods. 

\begin{table}[ht]
    \centering
    \begin{tabular}{cccccl}
    \hline
        &This work& \cite{Liu_etal_2021_BioAIP} & \cite{Bauer_etal_2019_ECGDynap} & \cite{Yan_etal_2021_ECGSNN} SNN \\
     \hline
       Power & $0.53\,$\textmu W & $48.6\,$\textmu W & $516.1\,$\textmu W & $64\,$mW \\
     \hline  
        Memory Footprint & 256~b & 73~kb & 64~kb & (NA)\\
    \hline
    \end{tabular}
    \caption{Energy and power consumption comparison with the state of the art.}
    \label{tab:power}
\end{table}

\section{Conclusions}

We have introduced an RRAM-aware dendritic architecture, which is empowered by delays, and as a result, can introduce temporal richness to a feedforward network that can classify a sensory processing task with up to 100x less power consumption and less than 100x in memory footprint compared to recurrent networks. The power benefits are thanks to the delays which keep the temporal information of the data in a passive fashion, without the need for active storage of data through recurrency. 


\begin{acks}
This work is supported by H2020 MeM-Scales project (871371), SNSF Starting Grant Project UNITE (TMSGI2-211461), and European Research Council consolidator grant DIVERSE (101043854).

\end{acks}


\bibliography{biblio,biblioncs}

\begin{thebibliography}{10}

\bibitem{paugam_etal_2012_SNN}
H{\'e}lene Paugam-Moisy and Sander~M Bohte.
\newblock Computing with spiking neuron networks.
\newblock {\em Handbook of natural computing}, 1:1--47, 2012.

\bibitem{pagkalos_etal2023_dendrify}
Michalis Pagkalos, Spyridon Chavlis, and Panayiota Poirazi.
\newblock Introducing the dendrify framework for incorporating dendrites to
  spiking neural networks.
\newblock {\em Nature Communications}, 14(1):131, 2023.

\bibitem{yin_etal2021_SRNNAdapt}
Bojian Yin, Federico Corradi, and Sander~M Boht{\'e}.
\newblock Accurate and efficient time-domain classification with adaptive
  spiking recurrent neural networks.
\newblock {\em Nature Machine Intelligence}, 3(10):905--913, 2021.

\bibitem{Wang_Liu10}
Y.~Wang and S.-C. Liu.
\newblock Multilayer processing of spatiotemporal spike patterns in a neuron
  with active dendrites.
\newblock {\em Neural Computation}, 8:2086--2112, 2010.

\bibitem{esmanhotto_etal_2022}
Eduardo Esmanhotto, Tifenn Hirtzlin, Djohan Bonnet, Niccolo Castellani,
  Jean-Michel Portal, Damien Querlioz, and Elisa Vianello.
\newblock Experimental demonstration of multilevel resistive random access
  memory programming for up to two months stable neural networks inference
  accuracy.
\newblock {\em Advanced Intelligent Systems}, 4(11):2200145, 2022.

\bibitem{Dalgaty_etal19}
T.~Dalgaty, M.~Payvand, B.~De~Salvo, J.~Casaz, G.~Lama, E.~Nowak, G.~Indiveri,
  and E.~Vianello.
\newblock Hybrid {CMOS-RRAM} neurons with intrinsic plasticity.
\newblock In {\em International Symposium on Circuits and Systems ({ISCAS}),
  2019}. IEEE, 2019.

\bibitem{esmanhotto_etal_2020}
E.~Esmanhotto, L.~Brunet, N.~Castellani, D.~Bonnet, T.~Dalgaty, L.~Grenouillet,
  D.~R.~B. Ly, C.~Cagli, C.~Vizioz, N.~Allouti, F.~Laulagnet, O.~Gully,
  N.~Bernard-Henriques, M.~Bocquet, G.~Molas, P.~Vivet, D.~Querlioz, JM.
  Portal, S.~Mitra, F.~Andrieu, C.~Fenouillet-Beranger, E.~Nowak, and
  E.~Vianello.
\newblock High-density 3d monolithically integrated multiple 1t1r
  multi-level-cell for neural networks.
\newblock In {\em 2020 IEEE International Electron Devices Meeting (IEDM)},
  pages 36.5.1--36.5.4, 2020.

\bibitem{Dedemirag_etal_2021_mixedprecision}
Yigit Demirag, Charlotte Frenkel, Melika Payvand, and Giacomo Indiveri.
\newblock Online training of spiking recurrent neural networks with
  phase-change memory synapses.
\newblock {\em CoRR}, abs/2108.01804, 2021.

\bibitem{legalo_etal2018_mixprec}
Manuel Le~Gallo, Abu Sebastian, Roland Mathis, Matteo Manica, Heiner Giefers,
  Tomas Tuma, Costas Bekas, Alessandro Curioni, and Evangelos Eleftheriou.
\newblock Mixed-precision in-memory computing.
\newblock {\em Nature Electronics}, 1(4):246--253, 2018.

\bibitem{moody_etal_2001_mitbih}
George~B Moody and Roger~G Mark.
\newblock The impact of the mit-bih arrhythmia database.
\newblock {\em IEEE engineering in medicine and biology magazine},
  20(3):45--50, 2001.

\bibitem{sharifshazileh_etal2021_eeg}
Mohammadali Sharifshazileh, Karla Burelo, Johannes Sarnthein, and Giacomo
  Indiveri.
\newblock An electronic neuromorphic system for real-time detection of high
  frequency oscillations (hfo) in intracranial eeg.
\newblock {\em Nature communications}, 12(1):3095, 2021.

\bibitem{barbot2022interplay}
Justine Barbot, Jean Coignus, Nicolas Vaxelaire, Catherine Carabasse, Olivier
  Glorieux, Messaoud Bedjaoui, Fran{\c{c}}ois Aussenac, Fran{\c{c}}ois Andrieu,
  Fran{\c{c}}ois Triozon, and Laurent Grenouillet.
\newblock Interplay between charge trapping and polarization switching in mfdm
  stacks evidenced by frequency-dependent measurements.
\newblock In {\em ESSCIRC 2022-IEEE 48th European Solid State Circuits
  Conference (ESSCIRC)}, pages 125--128. IEEE, 2022.

\bibitem{gutig_etal_2006_tempotron}
Robert G{\"u}tig and Haim Sompolinsky.
\newblock The tempotron: a neuron that learns spike timing--based decisions.
\newblock {\em Nature neuroscience}, 9(3):420--428, 2006.

\bibitem{Bauer_etal_2019_ECGDynap}
F.~C. Bauer, Muir, R.~Dylan, and G.~Indiveri.
\newblock Real-time ultra-low power ecg anomaly detection using an event-driven
  neuromorphic processor.
\newblock {\em IEEE Transactions on Biomedical Circuits and Systems},
  13(6):1575--1582, 2019.

\bibitem{Liu_etal_2021_BioAIP}
J.~Liu et~al.
\newblock 4.5 bioaip: A reconfigurable biomedical ai processor with adaptive
  learning for versatile intelligent health monitoring.
\newblock In {\em 2021 IEEE International Solid- State Circuits Conference
  (ISSCC)}, volume~64, pages 62--64, 2021.

\bibitem{Yan_etal_2021_ECGSNN}
Z.~Yan, J.~Zhou, and W.F. Wong.
\newblock Energy efficient ecg classification with spiking neural network.
\newblock {\em Biomedical Signal Processing and Control}, 63, 2021.

\end{thebibliography}
\bibliographystyle{unsrt}

\end{document}